\documentclass[10pt,twocolumn,letterpaper]{article}

\usepackage{cvpr}      

\usepackage{algpseudocode}
\usepackage{algorithm}
\usepackage{adjustbox}
\usepackage{multirow}
\usepackage{graphicx}
\usepackage{tabularx}
\usepackage{physics}
\usepackage[accsupp]{axessibility}

\algrenewcommand\algorithmicrequire{\textbf{Input:}}
\algrenewcommand\algorithmicensure{\textbf{Output:}}

\algnewcommand\algorithmicparams{\textbf{Parameters:}}
\algnewcommand\Params{\item[\algorithmicparams]}

\usepackage[dvipsnames]{xcolor}

\definecolor{cvprblue}{rgb}{0.21,0.49,0.74}
\usepackage[pagebackref,breaklinks,colorlinks,citecolor=cvprblue]{hyperref}

\def\paperID{31} 
\def\confName{CVPR}
\def\confYear{2024}

\title{Multimodal Attack Detection for Action Recognition Models}

\author{Furkan Mumcu, Yasin Yilmaz\\
University of South Florida\\
4202 E Fowler Ave, Tampa, FL 33620\\
{\tt\small \{furkan, yasiny\}@usf.edu}
}

\begin{document}
\maketitle

\begin{abstract}
Adversarial machine learning attacks on video action recognition models is a growing research area and many effective attacks were introduced in recent years. These attacks show that action recognition models can be breached in many ways. Hence using these models in practice raises significant security concerns. However, there are very few works which focus on defending against or detecting attacks. In this work, we propose a novel universal detection method which is compatible with any action recognition model. In our extensive experiments, we show that our method consistently detects various attacks against different target models with high true positive rates while satisfying very low false positive rates. Tested against four state-of-the-art attacks targeting four action recognition models, the proposed detector achieves an average AUC of 0.911 over 16 test cases while the best performance achieved by the existing detectors is 0.645 average AUC. This 41.2\% improvement is enabled by the robustness of the proposed detector to varying attack methods and target models. The lowest AUC achieved by our detector across the 16 test cases is 0.837 while the competing detector's performance drops as low as 0.211. We also show that the proposed detector is robust to varying attack strengths.  
In addition, we analyze our method's real-time performance with different hardware setups to demonstrate its potential as a practical defense mechanism. 



\end{abstract}    
\section{Introduction}
\label{sec:intro}

Adversarial machine learning attacks have been a very popular research topic since the introduction of the Fast Gradient Sign Method (FGSM) \cite{fgsm}. While the initial research on these attacks focused on image classification models, many recent works showed that adversarial attacks are effective against video understanding models including action recognition \cite{wei2020heuristic,yan2020sparse,li2021adversarial}, object detection \cite{lee2019physical,zhao2019seeing,jia2020fooling}, and anomaly detection \cite{mumcu2022adversarial}. Different architectural types such as convolutional neural networks (CNNs) and vision transformers (ViTs) are shown to be vulnerable against adversarial attacks \cite{wei2022towards, mumcu2023sequential}. 

Increasing numbers of successful adversarial attacks in a broad range of applications against various architectures raise real-world security concerns. However, compared to the attacks, defense or detection mechanisms against adversarial attacks have not been studied widely. 
Although there are several defense methods proposed for image recognition models, there is very limited work on defending against attacks targeting video recognition models. In this paper, we propose a \emph{Vision-Language Attack Detection (VLAD)} mechanism, which is the first vision-language based detection method for action recognition models.

Adversarial attacks are achieved by creating input data with perturbations which are not obvious to human eye, but result in errors for the machine learning algorithms. For images, this perturbation can be applied to the whole input \cite{pgd, fgsm} or a specific part of the input in the form of pixels or patches \cite{pixel2}. 
For videos, in addition to the spatial domain, attackers also have the opportunity to make perturbations in the temporal domain, which makes defending against video attacks a more challenging task. Compared to the image models, there are much less defense methods for action recognition models.
Advit \cite{advit} is an adversarial frame detection method which proposes using optical flow to generate pseudo frames and then evaluating the classifier output consistency between the original input frames and the pseudo frames. Shuffle \cite{shuffle} tries to defend against adversarial attacks by shuffling the input frames randomly before feeding into the action recognition model. In our extensive experiments we show that neither of these two methods perform well under many attack scenarios.  

Attacks against videos can be achieved in various ways due to the temporal nature of videos. In addition to possible perturbations in one frame (e.g., pixel, patch, all frame), one might combine these methods in many ways along the temporal dimension. Therefore, we believe detecting adversarial attacks against videos can be accomplished by observing the inputs and the model outputs with a separate-modality subsystem. 

In our approach, we use a vision-language model (VLM) as an observing subsystem. VLMs are gaining popularity with the introduction of CLIP \cite{clip} and are being used for many other applications such as video action recognition \cite{actionclp1,actionclp4}, object detection \cite{objclp1} and video anomaly detection \cite{anmlyclp1}. VLM's multimodal processing of visual and language features provide additional information space for detecting attacks based on only visual features. Leveraging the capability of VLMs to make connections between texts and images, our method utilizes the context of video by computing the similarity scores between the video frames and the action class labels. Then, it decides if the video is adversarial or clean based on the consistency/inconsistency between the predictions by the action recognition model and the similarity scores obtained by VLM. 

Our contributions can be summarized as follows:

\begin{itemize}
\item We propose 
a universal attack detection method which can easily work with any action recognition model.
\item To the best of our knowledge, this is the first method that leverages a vision-language model for context awareness against adversarial machine learning attacks.
\item We benchmark our detection with extensive experiments and analyze its effectiveness compared to existing defense methods. Experimental results show that our method 
exhibits a 41.2\% performance improvement relative to the state-of-the-art detector and robustness to varying attack methods and strengths. 
\item 
The real-time operation of the proposed method is demonstrated with several GPUs. 
\end{itemize}

\begin{figure}[t]
  \centering
   \includegraphics[width=\linewidth]{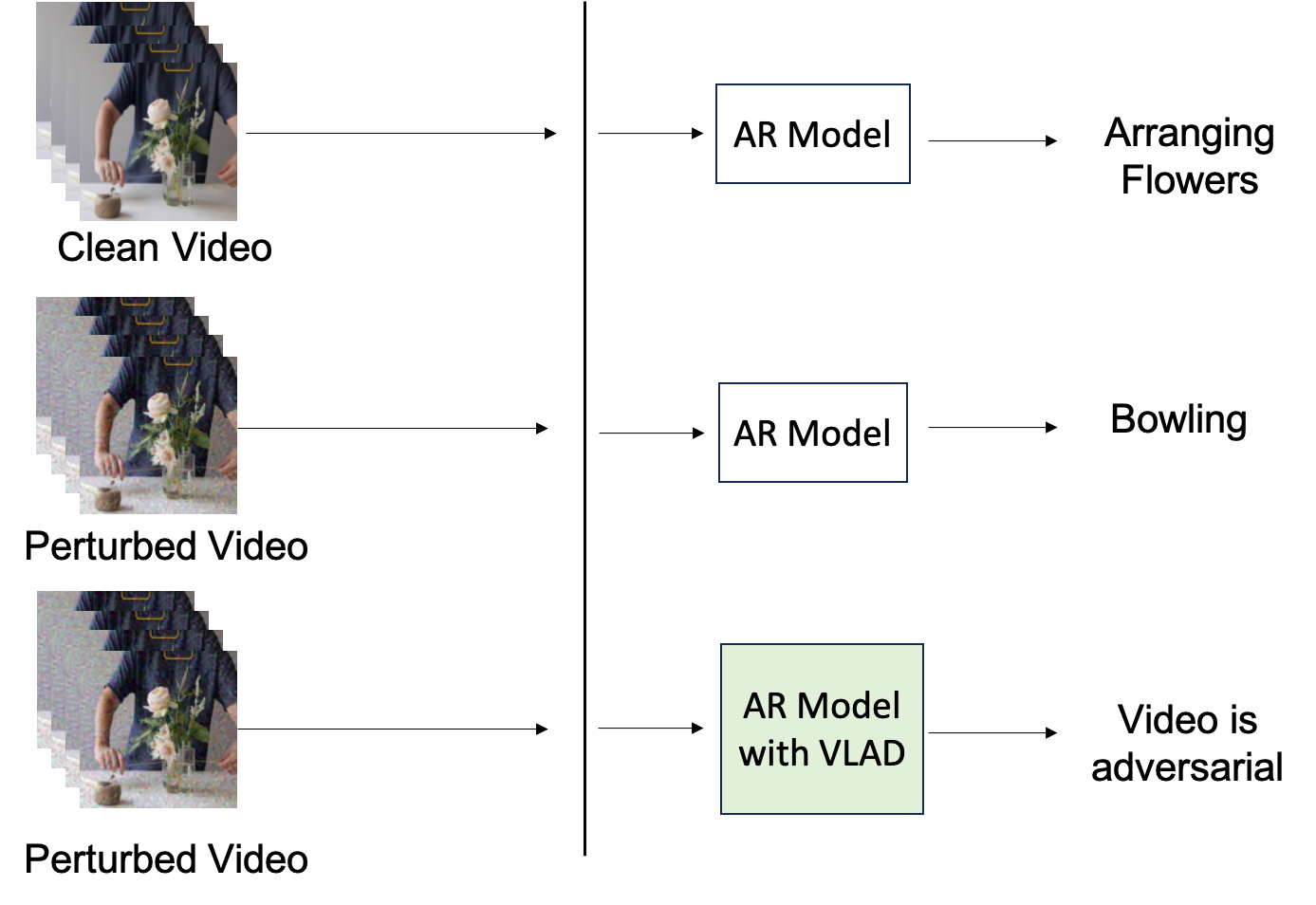}
   \caption{The proposed vision-language attack detection (VLAD) method cooperates with the action recognition (AR) model to detect adversarial videos.}
   \label{fig:problem}
   \vspace{-5mm}
\end{figure}
\section{Related Work}
\label{sec:related}

Adversarial machine learning attacks have been investigated for several years since the introduction of FGSM \cite{fgsm}. The first adversarial attacks were introduced for images under the black-box and white-box settings. In the white-box setting (e.g., FGSM \cite{fgsm}, CW \cite{cw}, PGD \cite{pgd}), an attacker has full access to the model including its architecture and parameters, while in the black-box setting (e.g., ZOO \cite{chen2017zoo}, NES \cite{ilyas2018black}, SPSA \cite{uesato2018adversarial}), the attacker has access only to the output labels or probabilities. 

After the increasing popularity of adversarial attacks on images, attacks against action recognition models have also been investigated broadly in recent years. Attacks like SparseAttack \cite{yan2020sparse}, Geo-Trap \cite{li2021adversarial} or HeuristicAttack \cite{wei2020heuristic} are query-based black-box attacks against video action recognition models. Although it is possible to adapt white-box image attacks to action recognition models, there are also several adversarial attacks designed for action recognition models. One Frame Attack (OFA) \cite{ofa} tries to attack models by perturbing one frame of the input. Flickering Attack (Flick) \cite{flick} tries to attack by changing the RGB stream of the input videos. In our experiments we test our defense method against white-box attacks since they provide more flexibility and effectiveness for adversaries, which make them more challenging to detect.

However, defense against adversarial attacks is not sufficiently investigated despite the increasing numbers and variety of attacks. Several defense methods have been developed for image models. \cite{yang2023advmask} aims to improve classification performance by adding an adversarial attack module and a data augmentation module to the model. \cite{yu2023improving} proposes a defense method where they analyze the common information between clean and perturbed data. \cite{niu2023defense} tries to remove perturbations with the help of adaptive compression and reconstruction. \cite{xie2017mitigating} implements random resizing to inputs to achieve robustness. \cite{dziugaite2016study} uses compression of JPEG images to avoid any possible perturbations. Some adversarial benchmark datasets, such as \cite{pintor2023imagenet}, were proposed to evaluate the robustness against adversarial attacks. 

There is not much research on defense against adversarial video attacks. Advit \cite{advit} is the first defense method introduced for videos. It 
generates pseudo frames using optical flow and evaluates the consistency between the outputs for original inputs and pseudo frames to detect attacks. Shuffle \cite{shuffle} is a recent defense method which aims to increase the robustness of action recognition models to attacks by randomly shuffling the input frames. 
They show that the predictions for clean videos are not significantly affected by shuffling while the adversarial effects of attacks are mitigated. Although Shuffle \cite{shuffle} is not an attack detection method, we show that it can be used for attack detection and compare our method with it. 

Vision language models are gaining popularity in recent years with the introduction of CLIP \cite{clip}. Many video understanding tasks benefit from VLMs including object detection \cite{objclp1, objclp2}, video action recognition \cite{actionclp1,actionclp4} and video anomaly detection \cite{anmlyclp1}. In our work, we present the first usage of VLMs for adversarial video detection. 
\section{Method}
\begin{figure*}[t]
  \centering
   \includegraphics[width=\linewidth]{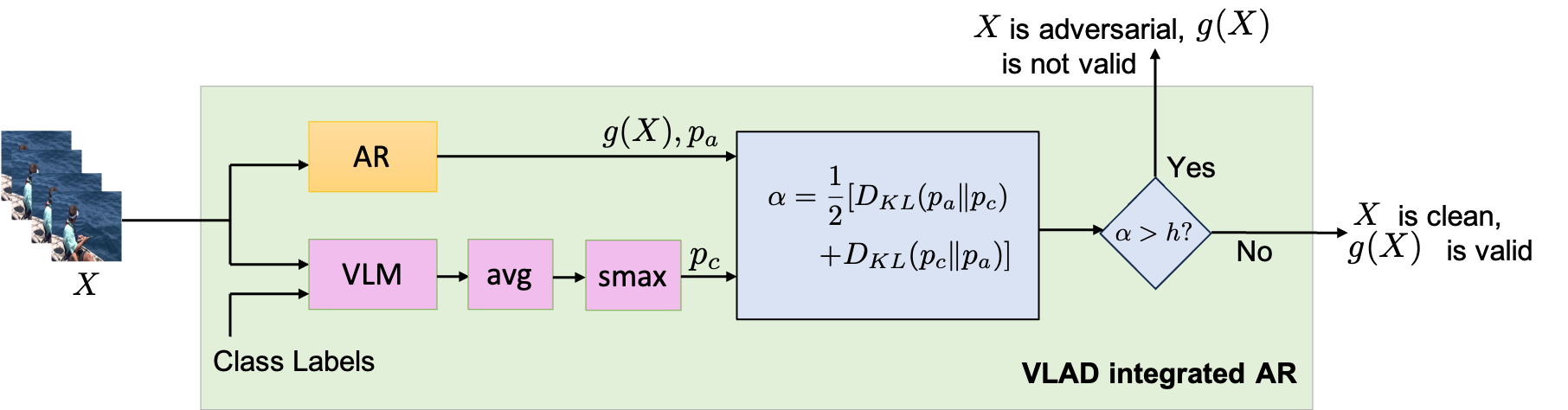}
   \caption{Overview of the proposed detection method. The predicted class $g(X)$ by the AR model is declared not valid if the attack detection statistic $\alpha$ is greater than the threshold $h$.}
   \label{fig:overview}
   \vspace{-5mm}
\end{figure*}

\subsection{Threat model} 

An action recognition (AR) model $g(\cdot)$ takes a video $ X  \in \mathbb{R}^{N \times H \times W \times C}$ as an input, which is a sequence of $N$ frames, each of which consists of $H \times W$ pixels and $C$ channels. Denoting the true label of the input with $y$, the true classification by the action recognition model is given by $g(X) = y$.
However, an adversary can attack the system by generating an adversarial version $X^{adv}$ of the input, which might be classified as $g(X^{adv}) = y'$, where $y \neq y'$. 
We consider the most challenging case for the detector, the white-box attack scenario, in which the attacker has full access to the AR model. The detector has also access to the predicted class probability vector $p_a = [p_1, p_2, p_3,...,p_M]$, where $M$ is the number of class labels.


\begin{algorithm}[t]
\caption{Vision-Language Attack Detection (VLAD)}\label{alg:cap}
\begin{algorithmic}[1]

\Require input video $X$, word vector $w$ containing class labels, action recognition model $g$, vision language model $vlm$, threshold $h$.

\Ensure detection result $d$.


\State action recognition model receives the input video, returns the classification probabilities
 $ p_a \gets g(X)$
\State $vlm$ calculates the similarity scores between the input video frames and word vector, 
$\{s_i\} \gets vlm(X, w) $
\State take the average of similarity scores over frames 
\newline $s \gets \frac{1}{N}\sum_{i=1}^N s_i$
\State apply softmax to similarity scores to get the context probabilities
 $p_c \gets \text{softmax}(s)$
\State Calculate detection score  
\newline $\alpha \gets \frac{1}{2} [D_{KL}(p_a||p_c) + D_{KL}(p_c||p_a)]$


\If {$\alpha > h $}
    \State $d \gets $ adversarial
\Else
    \State $d \gets $ not adversarial
\EndIf
\State \Return $d$

\end{algorithmic}
\end{algorithm}

\subsection{Proposed Method: VLAD}

There are many ways to generate successful video perturbation from changing only one frame to perturbing all frames. 
Due to the vastness of attack space and the typical obliviousness of  the action recognition models to the attack strategy, a defense mechanism should not use any bias regarding the attacks. Therefore, we propose a universal detection mechanism which does not rely on any assumption about the attack method or action recognition model, and hence can work with any model to detect a broad range of adversarial attacks (Fig. \ref{fig:problem}).

An overview of the proposed detection method is depicted in Fig. \ref{fig:overview}. 
In our detection mechanism, in parallel with the action recognition model, we apply a VLM to obtain context probabilities. We feed two inputs to the VLM to get the context similarity scores for each frame. Taking the input video frames $f = [f_1, f_2, \ldots, f_N]$ and the class labels $w = [w_1, w_2,\ldots,w_M]$ 
VLM outputs the similarity score matrix $S \in \mathbb{R}^{N \times M}$ for each input video $X$.



The similarity scores $s_i=[s_{i1},s_{i2},\ldots,s_{iM}]$ for each frame $i$ are averaged to obtain the similarity vector $s \in \mathbb{R}^M$ of the video:
\begin{equation}
\label{eq:similarity}
    s = \frac{1}{N} \sum_{i=1}^{N} s_i.
\end{equation}
Next, the softmax function is applied to the video similarity score to obtain the context probabilities:
\begin{equation}
    p_c = \text{softmax}(s) = [p_{c1}, p_{c2}, \ldots, p_{cM}].
\end{equation}




A final detection score $\alpha$ is calculated by getting the average of forward and reverse KL divergences between $p_a$ and $p_c$: 
\begin{equation}  
\label{eq:ds}
\alpha = \frac{1}{2} [D_{KL}(p_a||p_c) + D_{KL}(p_c||p_a)].
\end{equation}
The detection score $\alpha$ is expected to be low for clean inputs and high for adversarial inputs. A threshold $h$ is decided by calculating the detection scores for a set of clean videos, $\beta = [\alpha_1, \alpha_2, \ldots, \alpha_K]$, where $K$ is the number of clean videos and the scores in $\beta$ are sorted in ascending order. Threshold $h$ is selected as the $\theta$th percentile of the clean training scores:
\begin{equation}
\label{eq:thr}
h = \beta[\lfloor K \theta/100\rfloor],
\end{equation}
where $\lfloor\cdot\rfloor$ denotes the floor operator, and $\beta[i]$ denotes the $i$th element of $\beta$. 

After obtaining the detection score for an input, decision $d$ is made as follows:
\begin{equation}
d = 
\begin{cases}
    \text{$X$ is adversarial} & \text{if $\alpha > h $} \\
    \text{$X$ is not adversarial} & \text{if $\alpha \leq h $}, \\
\end{cases}
\end{equation}
where $\alpha$ is computed as in Eq. \eqref{eq:ds}. 


\subsection{Applicability and Implementation} 

Our proposed method is highly compatible with the existing action recognition models since it does not make any architectural changes to the existing models and it does not require any neural network training. VLAD only needs the classification probabilities from the action recognition model. Algorithm \ref{alg:cap} describes the overall workflow of VLAD. 
\footnote{\url{https://github.com/furkanmumcu/VLAD}} 
In our implementation we used CLIP \cite{clip} as VLM, but our method is compatible with any VLM. $N=32$ uniformly selected frames are used in Eq. \eqref{eq:similarity} to compute the similarity scores in VLM.

\section{Experiments}
\label{experiments}

\setlength{\extrarowheight}{3pt}

\begin{table*}[t]
\begin{adjustbox}
{width=.65\textwidth,center}
  \centering
  \begin{tabular}{c  c | c  c  c  c }
    
    &  & Advit \cite{advit} & Shuffle \cite{shuffle} & VLAD-1 & VLAD-2  \\
    \hline

            \multirow{4}{*}{PGD-v \cite{pgd}}
            &CSN \cite{csn} & 0.842 &	0.841	& \textbf{0.967}	& 0.937\\
		&SlowFast \cite{slowfast} &0.970	& 0.211 &	\textbf{0.977}	& 0.952\\
		&MVIT \cite{mvit} & 0.936 & 0.988	& \textbf{0.990} &	0.934 \\
		&X3D \cite{x3d} & 0.924 & 0.762 & \textbf{0.989} & 0.970\\ 
  \hline 

            \multirow{4}{*}{FGSM-v \cite{fgsm}}
            &CSN \cite{csn} & 0.429 & 0.620 &	0.824 &	\textbf{0.895}\\
		&SlowFast \cite{slowfast} &0.513	& 0.400 &	0.870 &	\textbf{0.933}\\
		&MVIT \cite{mvit} & 0.384	& \textbf{0.920} &	0.759 &	0.837 \\
		&X3D \cite{x3d} & 0.466 & 0.495 & 0.776 & \textbf{0.951} \\ 
  \hline
            \multirow{4}{*}{OFA \cite{ofa}}
            &CSN \cite{csn} & 0.370 &	0.694 &	0.680 &	\textbf{0.904}\\
		&SlowFast \cite{slowfast} & 0.451 &	0.495 & 0.797 &	\textbf{0.935}\\
		&MVIT \cite{mvit} & 0.570 & \textbf{0.907} &	0.719 &	0.876 \\
		&X3D \cite{x3d} & 0.581 & 0.505 &	0.572 &	\textbf{0.926} \\ 
  \hline
            \multirow{4}{*}{Flick \cite{flick}}
            &CSN \cite{csn} & 0.252 &	0.629 &	0.548 &	\textbf{0.883}\\
		&SlowFast \cite{slowfast} & 0.267 &	0.611 &	0.600 &	\textbf{0.878}\\
		&MVIT \cite{mvit} & 0.341 &	0.651 &	0.265 &	\textbf{0.870} \\
		&X3D \cite{x3d}& 0.540 &	0.596 &	0.455 &	\textbf{0.906} \\ 
  \hline
            & Average & 0.552 &	0.645 &	0.736 &	\textbf{0.911}\\

  \end{tabular}
  \end{adjustbox}
  \caption{Attack detection results (AUC) for defense methods Advit \cite{advit}, Shuffle \cite{shuffle}, VLAD-1, and VLAD-2, where the adversarial attacks (PGD-v \cite{pgd}, FGSM-v \cite{fgsm}, OFA \cite{ofa} and FLICK \cite{flick}) target action recognition models CSN \cite{csn}, SlowFast \cite{slowfast}, MVIT \cite{mvit}, and X3D \cite{x3d}. }
  \label{tab:main-results}
\vspace{-5mm}
\end{table*}

For performance evaluation, we compare our proposed detection mechanism with the existing defense methods. First, we report the detection performances against four adversarial attacks when they target four different action recognition models. Then, we examine the performance against a specific attack with different strength settings. 

\textbf{Dataset:} Kinetics-400 \cite{k400} is a popular and large-scale dataset for action recognition. A subset of Kinetics-400 is randomly selected for each target model from the videos that are correctly classified by the respective model. For each subset, the total number of the videos are between 7700 and 8000 and each class has at least 3, at most 20 instances. During the experiments, 80\% of the videos are used for training, which is the process of getting the detection scores from a clean set of videos to set the threshold $h$ as in Eq. \eqref{eq:thr}. 
An adversarial version of the remaining 20\% portion is generated with each adversarial attack, in a way that they cannot be correctly classified by the models. Then the adversarial set is used for evaluation along with the clean versions. Hence, the test set consists of equal number of clean and adversarial videos. 

\textbf{Target Models:} We use the following popular video action recognition models as
target in our experiments: MVIT \cite{mvit}, CSN \cite{csn}, X3D \cite{x3d} and SlowFast \cite{slowfast}. While  CSN \cite{csn}, X3D \cite{x3d} and SlowFast \cite{slowfast} are CNN based models, MVIT \cite{mvit} is a transformer based model.

\textbf{Adversarial attacks:} In our experiments we used four different adversarial attacks. Fast Gradient Sign Method (FGSM) \cite{fgsm} and and Projected Gradient Descent (PGD) \cite{pgd} are strong adversarial attacks that originally targets images. We adopted these methods for videos and generated FGSM-v and PGD-v respectively by taking the gradients for entire video and perturbing all frames. One Frame Attack (OFA) \cite{ofa} targets action recognition models by selecting and attacking a specific frame. Flickering Attack (Flick) \cite{flick} changes the RGB stream of the frames. For each attack, we used the attack settings that were originally proposed by their authors.

\textbf{Defense methods:} Advit \cite{advit} is the first defense method designed for videos. It aims to detect adversarial frames for semantic segmentation, object detection and pose estimation tasks; and adversarial videos for action recognition. It generates pseudo frames for the inputs using the optical flow and observes the consistency between the predicted class probabilities of the original frames and the pseudo frames. Similarly to our method, it also uses the KL divergence as the consistency measure.  

Shuffle \cite{shuffle} is a recent defense method that aims to make action recognition models robust to attacks. It tries to improve robustness by shuffling the frames of the input video. It is claimed that the prediction for clean videos are not affected by shuffling. Since Shuffle is not originally a detection method, we obtain a detection method from it by calculating the KL divergence between the action recognition model's class probabilities for the original video and shuffled video.

We consider two versions of our approach in which we use different representations of the predicted class probabilities $p_a$ from the action recognition model while calculating the KL divergence described in Eq. \eqref{eq:ds}. In the first version, VLAD-1, the original probabilities directly coming from the model are used, whereas in the second version, VLAD-2, one-hot-encoding of predicted class is used (i.e., probability of the predicted class set to 1 and all the others to 0). 

\textbf{Evaluation metric:} To evaluate the attack detection performance of defense methods, we report the commonly used the Area Under Curve (AUC) metric from the Receiver Operating Characteristic (ROC) curve, which shows the tradeoff between true positive rate (i.e., ratio of successfully detected adversarial videos to all adversarial videos) and false positive rate (i.e., ratio of false alarms to all clean videos).

\subsection{Comparison of Detection Methods}
\label{main_exp}

In Table \ref{tab:main-results}, we report the AUC scores for VLAD-1, VLAD-2, Shuffle, and Advit \cite{advit} against the PGD-v \cite{pgd}, FGSM-v \cite{fgsm}, OFA \cite{ofa}, and Flick \cite{flick} attakcs targeting 4 models. Since it is known that attacks might have different performances on different architectures \cite{mumcu2023sequential}, we selected both CNN-based (CSN, SlowFast, X3D) and transformer-based (MVIT) target models. We used the same parameter settings as reported in the original source for the attack and target models. 

Except the FGSM-v \cite{fgsm} and OFA \cite{ofa} attacks on MVIT \cite{mvit}, in all cases VLAD outperforms the existing defense methods. Even in those cases, the performance of VLAD is not low, 0.837 and 0.876 AUC, respectively, showing the robustness of VLAD to varying attack and target models. On the other hand, the existing methods cannot successfully detect attacks in several cases. Advit and Shuffle cannot exceed the random guess level of 0.5 AUC in 8 and 4 of the 16 scenarios, respectively. This significant difference in robustness results in a wide gap between the average AUC performances over all 16 scenarios: VLAD-2 outperforms Advit by 0.359 AUC (65\% relative performance improvement) and Shuffle by 0.266 (41.2\% relative performance improvement). 

While against PGD-v \cite{pgd}, VLAD-1 performs slightly better than VLAD-2, in all other attack-target combinations VLAD-2 achieves the best scores. This is due to PGD-v being the strongest attack among all, as shown in Table \ref{tab:attack-str}, which shows the mean and standard deviation of the probability for the wrongly predicted class after the attack. Without a defense mechanism, the AR models are very confidently fooled by the PGD-v attack. For the same reason, all defense methods successfully detect all PGD-v attacks, except for Shuffle with SlowFast. We investigate different attack strengths for PGD-v in Section \ref{sec:attac-str}. Conversely, the other attacks, especially the Flick attack, do not cause high confidence in AR models' wrong predictions. As a result, VLAD-2 considerably improves over VLAD-1 by amplifying the predicted probabilities and the KL divergence in detection. 

\begin{table}
  \centering
  \begin{adjustbox}{width=\columnwidth,center}
  \begin{tabular}{c | c | c | c  |c }
  
    & CSN \cite{csn} & SlowFast \cite{slowfast} & MVIT \cite{mvit} & X3D \cite{x3d} \\
    \hline
    PGD-v \cite{pgd} & 0.94 $\pm$ 0.1	& 0.92 $\pm$ 0.17 & 0.99 $\pm$ 0.1 &	0.96 $\pm$ 0.1 \\
    FGSM-v \cite{fgsm} & 0.38 $\pm$ 0.24 &	0.23 $\pm$ 0.24 &	0.58 $\pm$ 0.28 &	0.66 $\pm$ 0.28 \\
    OFA \cite{ofa} & 0.42 $\pm$ 0.22	& 0.12 $\pm$ 0.14 &	0.73 $\pm$ 0.26 &	0.67 $\pm$ 0.26 \\
    Flick \cite{flick} & 0.06 $\pm$ 0.03 &	0.12 $\pm$ 0.14 &	0.39 $\pm$ 0.2 &	0.37 $\pm$ 0.22

  \end{tabular}
  \end{adjustbox}
  \caption{Mean and standard deviation of wrongly predicted class probabilities of target models CSN \cite{csn}, SlowFast \cite{slowfast}, MVIT \cite{mvit} and X3D \cite{x3d} when they are attacked with PGD-v \cite{pgd}, FGSM-v \cite{fgsm}, OFA \cite{ofa} and Flick \cite{flick}.  }
  \label{tab:attack-str}
  \vspace{-3mm}
\end{table}

\subsection{Comparison of Attacks}

\begin{table}
  \centering
  \begin{tabular}{c c c c }
    PGD-v \cite{pgd} & FGSM-v \cite{fgsm} & OFA \cite{ofa} & Flick \cite{flick} \\
    \hline
    0.887 & 0.692 & 0.686 & 0.581

  \end{tabular}
  \caption{Average AUC performance of all defense methods for all AR models. Smaller values indicate more effective attack.}
  \label{tab:attacks}
  \vspace{-4mm}
\end{table}

Table \ref{tab:attacks} presents the AUC values averaged over the four defense methods for the four AR models attacked by each method. According to the results, with the parameters given in the original papers, Flick is the most stealthy attack allowing only 0.581 average AUC while PGD-v is the least stealthy one having been detected by the considered defense mechanisms most of the time with 0.887 average AUC. These results are consistent with the wrong confidence levels caused by attacks reported in Table \ref{tab:attack-str}. PGD-v, being a dominant attack, can be the most cunning one in the absence of a defense mechanism, whereas other attacks are able to stay more stealthy even in the presence of defense by causing more uncertainty in the confidence levels of predicted class. 

\subsection{Comparison of AR Models}

\begin{table}
  \centering
  \begin{tabular}{c c c c }
    CSN \cite{csn} & SlowFast \cite{slowfast} & MVIT \cite{mvit} & X3D \cite{x3d} \\
    \hline
    0.707 & 0.678 & 0.746 & 0.713

  \end{tabular}
  \caption{Average AUC performance of all defense methods against all attacks. Larger values indicate more defensible model.}
  \label{tab:models}
  \vspace{-5mm}
\end{table}

Table \ref{tab:models} shows the AUC values averaged over all defense methods and all attacks targeting each AR model. 
We observe that there is no significant performance changes due to the architectural differences of target models. 
According to the results, MVIT is the most defensible model while SlowFast is the least defensible one.

\subsection{Robustness to varying attack strength}
\label{sec:attac-str}

\begin{figure*}[t]
  \centering
   \includegraphics[width=\linewidth]{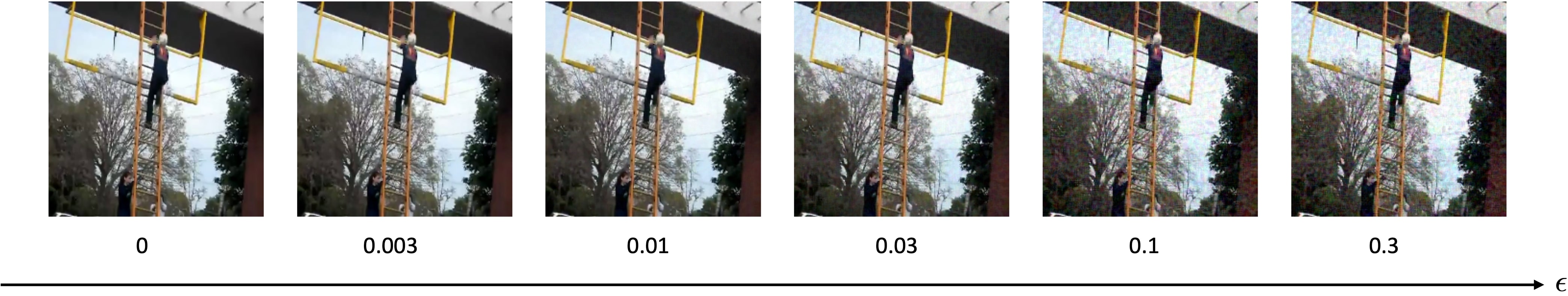}
   \caption{Effects of different attack strength values on a frame. The attack strength parameter $\epsilon$ is increased in the direction of the arrow. In the first sample, there is no attack. Samples are generated with adversarial attack PGD-v \cite{pgd} and target model MVIT \cite{mvit}.}
   \label{fig:str-examples}
   \vspace{-5mm}
\end{figure*}

\begin{figure*}[t]
  \centering
  \minipage{0.5\textwidth}
   \centering
   \includegraphics[width=.8\linewidth]{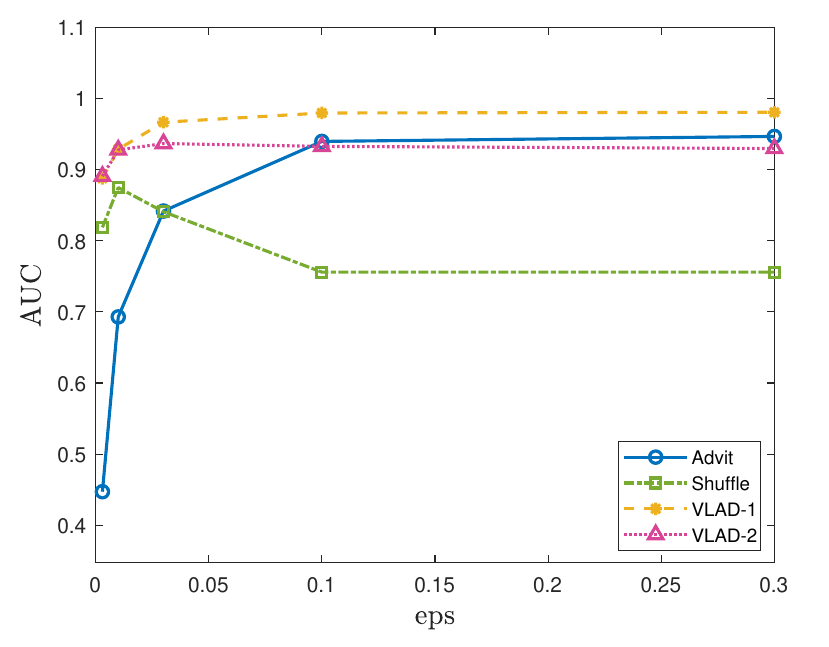}
   \subcaption{CSN}
   \endminipage \hfill
   \minipage{0.5\textwidth}
   \centering
   \includegraphics[width=.8\linewidth]{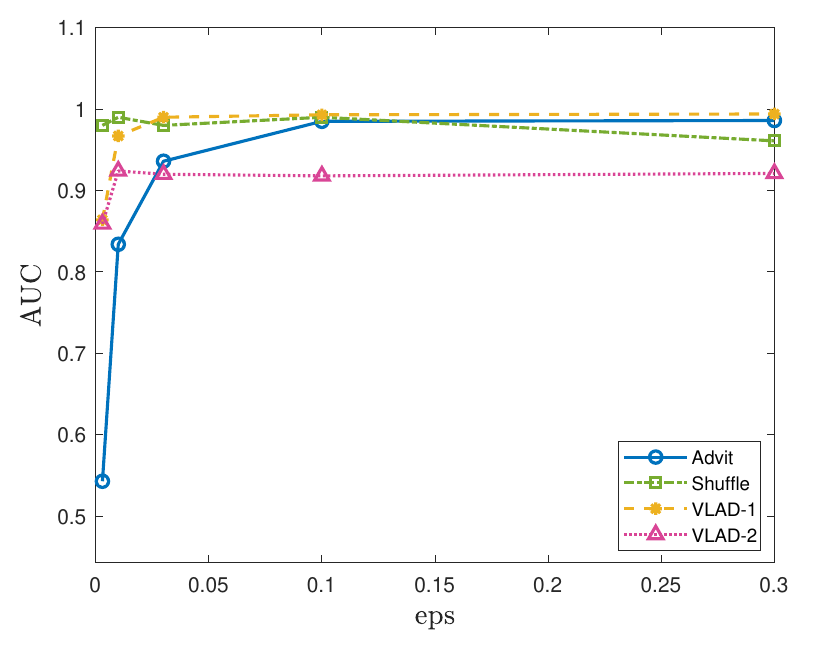}
   \subcaption{MVIT}
   \endminipage \hfill
   \minipage{0.5\textwidth}
   \centering
   \includegraphics[width=.8\linewidth]{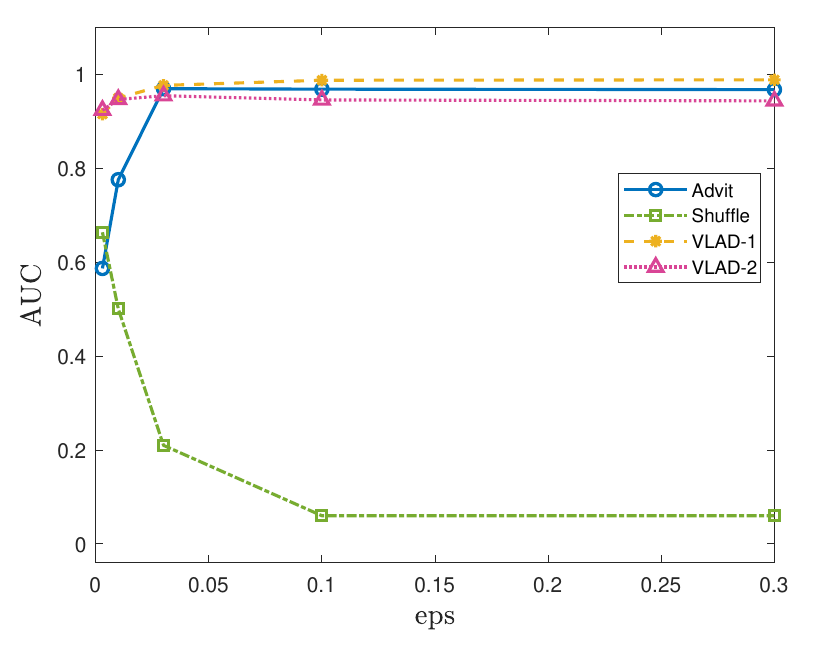}
   \subcaption{SlowFast}
   \endminipage
    \minipage{0.5\textwidth}
   \centering
   \includegraphics[width=.8\linewidth]{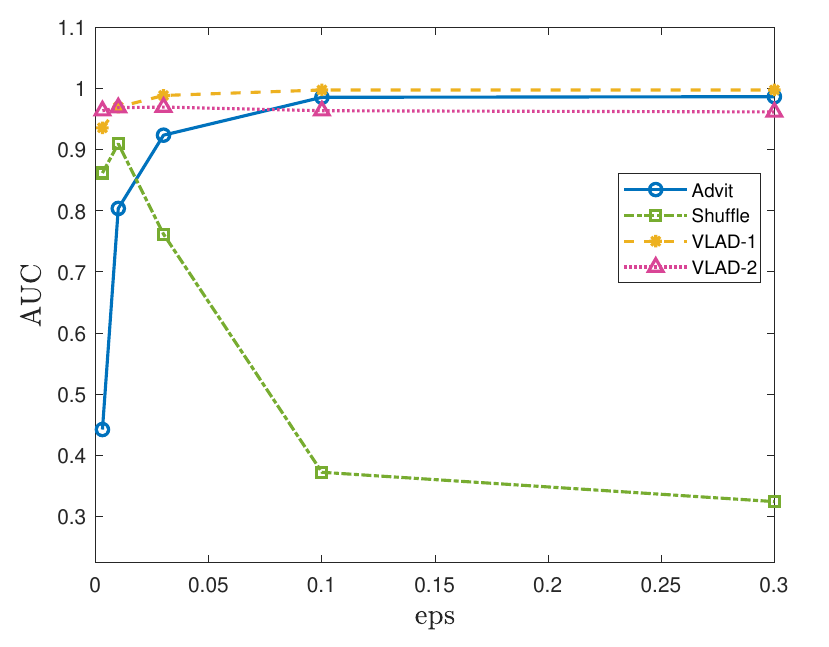}
   \subcaption{X3D}
   \endminipage
   \caption{Change in AUC as a function of PGD-v attack strength parameter $\epsilon$. Advit \cite{advit} loses performance against stealthy attacks (low $\epsilon$) while Shuffle \cite{shuffle} cannot detect strong attacks. The proposed models, VLAD-1 and VLAD-2, are robust to changes in the attack strength.}
   \label{fig:attack-strengths}
   \vspace{-5mm}
\end{figure*}

Attack strength is an important aspect in adversarial machine learning. For the experiments in Section \ref{main_exp}, we used the default attacks settings that are proposed by the authors and we noticed that their effects on target models vary. 
According to Table \ref{tab:main-results}, all defense methods performs relatively well against PGD-v \cite{pgd} compared to other attacks. Therefore, we analyzed PGD-v \cite{pgd} with different strengths. In the original implementation of PGD \cite{pgd}, attack strength parameter $\epsilon$ is set to 0.03. In addition to this value, we also generated attacks by setting $\epsilon$ to 0.003, 0.01, 0.1 and 0.3. 
As illustrated in Figure \ref{fig:str-examples}, we chose $\epsilon=0.003$ as the lower bound to ensure the attacks are still able to fool the AR models and  $\epsilon=0.3$ as the upper bound to prevent the perturbations from becoming too visible to human eye.

In Figure \ref{fig:attack-strengths}, the changing AUC values due to different $\epsilon$ values are presented for each target model. It is seen that, except for the attack on MVIT \cite{mvit}, Shuffle \cite{shuffle} does not perform well against strong attacks. Especially on Slowfast \cite{slowfast} AUC drops to 0.06 when the attack strength is 0.1 and 0.3. On contrary, Advit \cite{advit} does not perform well against weak attacks.  For all of the target models, both of our methods, VLAD-1 and VLAD-2, are not affected by the strength of the PGD-v attack.

\section{Ablation Study}

In this section, first we investigate other possible approaches for our proposed detection method VLAD. Then, we analyze VLAD's real-time performance and discuss its real-world applicability.

\subsection{VLAD with different score calculations}

In Eq. \eqref{eq:ds}, we use KL divergence to compute the detection score. Here we also consider 4 more ways to compute the detection score. 

In the first approach, we first normalize the predicted class probabilities $p_a$ from the AR model and $p_c$ from the VLM with their L2 norms:
\begin{equation}
    \Tilde{p}_a = \frac{p_a}{\|p_a\|}, ~~\Tilde{p}_c = \frac{p_c}{\|p_c\|}.
\end{equation}
Then, the detection score is computed using the absolute difference between their maximum probabilities:
\begin{equation}  
\label{eq:ds1}
\alpha_1 =  |\Tilde{p}_a[\text{max}]-\Tilde{p}_c[\text{max}]|,
\end{equation}
where $[\text{max}]$ denotes the maximum element of vector. 

In the second approach, we directly took the absolute difference between the predicted classes' probabilities without any normalization:
\begin{equation}  
\label{eq:ds2}
\alpha_2 = |p_a[\text{max}]-p_c[\text{max}]|.
\end{equation}

In the third approach, again we apply L2 normalization to both $p_a$ and $p_c$, but instead of using the maximum probabilities, we take the sum of absolute differences: 
\begin{equation}  
\label{eq:ds3}
\alpha_3 =  \sum |\Tilde{p}_a  - \tilde{p}_c |.
\end{equation}

Lastly, we directly take the sum of absolute difference without normalization:
\begin{equation}  
\label{eq:ds4}
\alpha_4 =  \sum |p_a - p_c|.
\end{equation}

We investigate the performances of the four score calculations by using a small subset of Kinetics-400, which has 10 distinct classes and 50 instances for each class, totaling 500 videos. We attack MVIT with PGD-v and FGSM-v, then report the average AUCs for different detection score calculations.

Figure \ref{tab:alt_score} shows the average AUC results for each score calculation approach described in this section. $\alpha_{VLAD-1}$ is the version proposed in equation \ref{eq:ds}, where we use the original probability predictions of the action recognition model.$\alpha_{VLAD-2}$ is the version in which we set the predicted class probability to 1 and the remaining probabilities to 0. These results guided us to use $\alpha_{VLAD-1}$ and $\alpha_{VLAD-2}$ in the final version of our proposed method.


\begin{table}
  \centering
  \begin{tabular}{c c c c c c }
    $\alpha_1$ & $\alpha_2$ & $\alpha_2$ & $\alpha_4$ & $\alpha_{VLAD-1}$ & $\alpha_{VLAD-2}$ \\
    \hline
    0.84 & 0.65 & 0.677 & 0.83 & 0.92 & 0.99

  \end{tabular}
  \caption{AUC results with different score calculations.}
  \label{tab:alt_score}
  \vspace{-4mm}
\end{table}


\subsection{Real-time performance}

\begin{figure}[t]
  \centering
   \includegraphics[width=1\columnwidth]{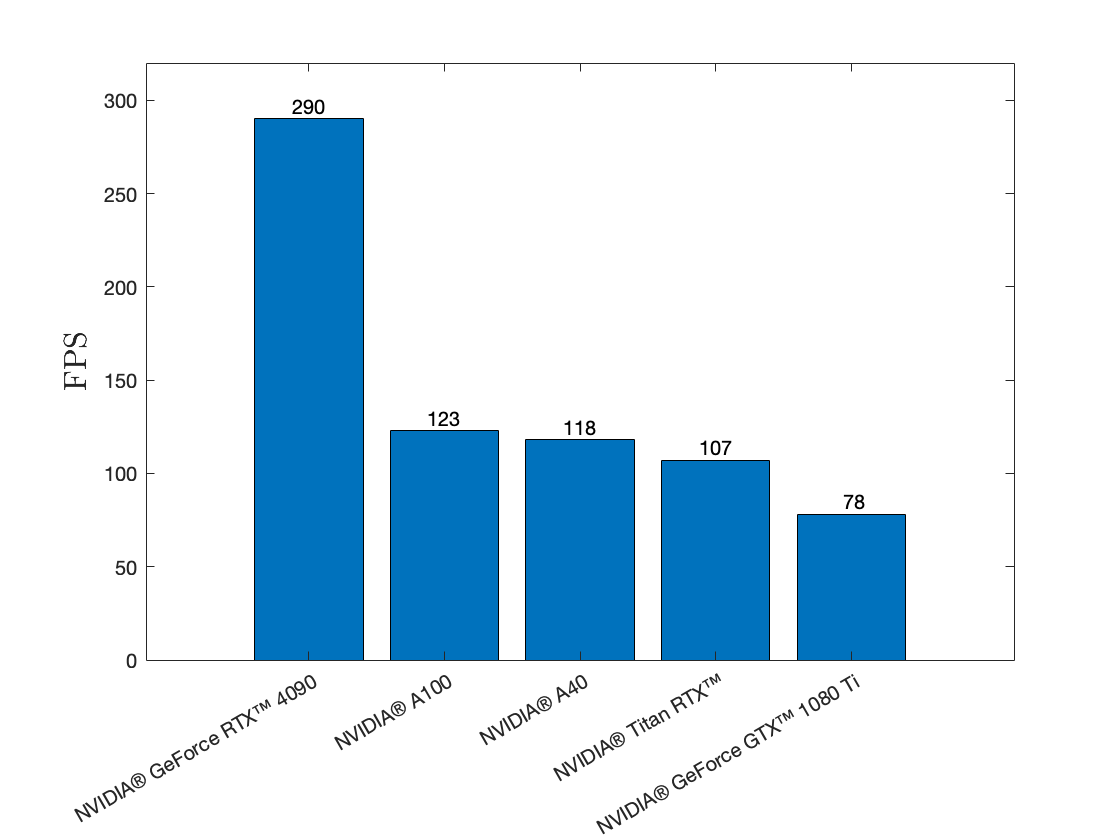}
   \caption{Frames per second (FPS) performance of proposed method on different GPUs. It provides real-time performace even on obsolete GPUs.}
   \label{fig:gpus}
   \vspace{-5mm}
\end{figure}

Real-time performance is a crucial aspect for an attack detection method. A detection mechanism needs to run always along with the action recognition model for timely detection of attacks. We benchmarked our detection mechanism with different hardware. We noticed that changing CPU does not affect the performance, therefore we tested our method with five different GPUs, namely NVIDIA® GeForce RTX™ 4090, NVIDIA® A100, NVIDIA® A40, NVIDIA® Titan RTX™, and NVIDIA® GeForce GTX™ 1080 Ti. In Figure \ref{fig:gpus}, we report the total number of frames per second (FPS) that can be processed by our detection mechanism. While NVIDIA® GeForce RTX™ 4090 has the best performance with 290 FPS, NVIDIA® GeForce GTX™ 1080 Ti has the worst with 78 FPS. These results show that even with an obsolete GPU our method can process in real-time streaming videos which are typically 30 FPS.

\section{Limitations and Future Work}

This study was limited to white-box attacks. In a future work, we plan to investigate detection against black-box attacks. Since in theory white-box attacks are more capable than black-box attacks, we believe VLAD can also successfully detect black-box attacks. With further improvements to our method, we aim to detect black-box attacks during their query-sending phase.

With the introduction of VLAD, we believe that an important portion of adversarial attacks developed for action recognition models can be detected. However, a future attack which can also fool context-aware VLM can be successful against our detection mechanism. 

Extension of this study to other video understanding systems (e.g., real time object detection and video anomaly detection systems) is a natural future research direction. 

\section{Conclusion}

The increasing number of successful attacks against action recognition (AR) models in recent years raises real-world security concerns. With this motivation, in this paper, we proposed a novel Vision-Language Attack Detection (VLAD) mechanism, which is the first vision-language model based attack detection method. The proposed VLAD method is a universal detector in the sense that it does not rely on any knowledge of attack or AR model. 


To benchmark our performance and compare it with existing defense methods, we have conducted extensive experiments with different attack methods and AR models. Experimental results show that VLAD consistently outperforms the existing methods by a wide margin over a wide range of attack and AR model scenarios. While the existing methods perform well in some scenarios, their performance drops below the random guess level of 0.5 AUC in several scenarios. The lowest performance of the proposed VLAD method in all scenarios was found to be 0.837. The average AUC value of VLAD over all scenarios is 0.911, which presents a 41.2\% improvement relative to the state-of-the-art result of 0.645 average AUC. VLAD is also robust to different attack strengths. While the performance of existing methods deteriorate against either stealthy or strong attacks, VLAD's performance remain steady against a wide range of attacks. Finally, we analyzed its real-time performance and showed that it can perform real-time detection even with an obsolete GPU. 



\textbf{Acknowledgements.} This work was supported by US National Science Foundation under the grant \#2040572.

{
    \small
    \bibliographystyle{ieeenat_fullname}
    \bibliography{main}
}


\end{document}